\documentclass[12pt]{article}
\pdfoutput=1
\usepackage[utf8]{inputenc}
\usepackage{amssymb}
\usepackage{amsmath}
\usepackage{amsthm}
\usepackage{amsfonts}
\usepackage[pdftex]{graphicx}
\usepackage{geometry}
\usepackage{braket}
\usepackage{enumerate}
\usepackage{comment}
\usepackage{subcaption}

\allowdisplaybreaks

\usepackage[table,usenames,dvipsnames]{xcolor}

\usepackage{setspace}
\usepackage[in]{fullpage}
\usepackage{hyperref}
\usepackage{array}
\usepackage{cancel}
\usepackage{wrapfig}
\usepackage[font=small,labelfont=bf]{caption}
\usepackage{pifont}

\usepackage[dvipsnames]{xcolor}
\usepackage{mathtools}

\usepackage{color}

\usepackage{tikz}
\usetikzlibrary{arrows,shapes}
\usetikzlibrary{trees}
\usetikzlibrary{matrix,arrows} 				% For commutative diagram
\usetikzlibrary{positioning}				% For "above of=" commands
\usetikzlibrary{calc,through}				% For coordinates
\usetikzlibrary{decorations.pathreplacing}  % For curly braces
\usepackage{pgffor}							% For repeating patterns

\usetikzlibrary{decorations.pathmorphing}	% For Feynman Diagrams
\usetikzlibrary{decorations.markings}
\tikzset{
    vector/.style={decorate, decoration={snake}, draw},
	provector/.style={decorate, decoration={snake,amplitude=2.5pt}, draw},
	antivector/.style={decorate, decoration={snake,amplitude=-2.5pt}, draw},
    fermion/.style={draw=black, postaction={decorate},
        decoration={markings,mark=at position .55 with {\arrow[draw=black]{>}}}},
    fermionbar/.style={draw=black, postaction={decorate},
        decoration={markings,mark=at position .55 with {\arrow[draw=black]{<}}}},
    fermionnoarrow/.style={draw=black},
    gluon/.style={decorate, draw=black,
        decoration={coil,amplitude=4pt, segment length=5pt}},
    scalar/.style={dashed,draw=black, postaction={decorate},
        decoration={markings,mark=at position .55 with {\arrow[draw=black]{>}}}},
    scalarbar/.style={dashed,draw=black, postaction={decorate},
        decoration={markings,mark=at position .55 with {\arrow[draw=black]{<}}}},
    scalarnoarrow/.style={dashed,draw=black},
    electron/.style={draw=black, postaction={decorate},
        decoration={markings,mark=at position .55 with {\arrow[draw=black]{>}}}},
	bigvector/.style={decorate, decoration={snake,amplitude=4pt}, draw},
}

\tikzstyle{block} = [draw, rectangle, 
    minimum height=3em, minimum width=6em]

\numberwithin{equation}{section}

\hypersetup{
	colorlinks=true,
	linktoc=page,
	citecolor=Blue,
	linkcolor=Blue,
	urlcolor=Blue} 

\makeatletter % Need for anything that contains an @ command 
\renewcommand{\maketitle} % Redefine maketitle to conserve space
{ \begingroup \begin{center} \large {\bf \@title}
		\vskip 5pt \large \@author \\ \vskip 5pt \@date \end{center}
	\vskip 5pt \endgroup \setcounter{footnote}{0} }
\makeatother % End of region containing @ commands

\newcommand{\comments}[1]{}

\newcommand{\be}{\begin{equation}}
\newcommand{\ee}{\end{equation}}

\def\beqa{\begin{eqnarray}}
\def\eeqa{\end{eqnarray}}
\def\beq{\begin{equation}}
\def\eeq{\end{equation}}

\def\one{\mbox{1 \kern-.59em {\rm l}}}

 \def\C_B{{\cal B}} 
  \def\C_F{{\cal F}}
  
  \def\cL{{\cal L}}
  \def\cO{{\cal O}}

\def\uno{\mbox{1 \kern-.59em {\rm l}}}

\def\one{1\!\!1\,\,}

\def\bcomment#1{}
\def\eps{\epsilon}

\usepackage{slashed}
\usepackage{caption}

\usepackage[nottoc,notlot,notlof]{tocbibind}
\usepackage[nosort]{cite}
\usepackage{color}

\usepackage{parskip}
\graphicspath{{./Images/}}

\long\def\symbolfootnote[#1]#2{\begingroup%
	\def\thefootnote{\fnsymbol{footnote}}\footnote[#1]{#2}\endgroup}

\begin{document}
	
	\begin{flushright}
		QMUL-PH-19-09\\
		SAGEX-19-05\\
		%HU-EP-18/25
	\end{flushright}
	
	\vspace{15pt} % was 20

	\begin{center}
		
		{\Large \bf  On higher-derivative effects on the }  \\
		\vspace{0.3 cm} {\Large \bf  gravitational potential and particle bending}

		%
		%\vspace{45pt} % was 45
		\vspace{25pt}
		%\end{center}

		%{\mbox {\bf   A.~Brandhuber$^{a,b,\S}$, 
		%		M.~Kostaci\'{n}ska$^{a,\S}$, 
		%				B.~Penante$^{c,\star}$ \!and  %
		%		G.~Travaglini$^{a,b,d,\S}$}}%
		
		{\mbox {\bf  Andreas~Brandhuber 
				%}} \\ \vspace{0.2cm}
				%{\mbox{\bf
				 and  %
				Gabriele~Travaglini$^{\alpha^\prime}$}}%
	%	\symbolfootnote[4]{
	%	{\tt  \{ \tt \!\!\!a.brandhuber, m.m.kostacinska, g.travaglini\}@qmul.ac.uk}
	%	}

		\vspace{0.5cm}
		
		\begin{center}
			{\small \em
				%\begin{itemize}
				%	\item[\ \ \ \ \ \ ]
					%\begin{flushleft}
						Centre for Research in String Theory\\
						School of Physics and Astronomy\\
						Queen Mary University of London\\
						Mile End Road, London E1 4NS, United Kingdom
				%	\end{flushleft}

				%	\item[\ \ \ \ \ \ $^b$]
				%	Institut f\"{u}r Physik und IRIS Adlershof\\
				%	Humboldt-Universit\"{a}t zu Berlin\\
				%	Zum Gro{\ss}en Windkanal 6, 12489 Berlin, Germany

				%\end{itemize}
			}
		\end{center}

		%\vspace{-8pt}

		\vspace{15pt}  %was 40 

{\bf Abstract}
	\end{center}
	
	\vspace{0.3cm} 
	
\noindent

\noindent
Using modern amplitude techniques we compute the leading classical and quantum corrections  to the  gravitational potential between two massive scalars induced by adding   cubic  terms to  Einstein gravity.  We then study the 
scattering of massless scalars, photons and gravitons off a heavy scalar in the presence of the same $R^3$ deformations, and determine the bending angle in the three cases from the non-analytic component of the  scattering amplitude. Similarly to the Einstein-Hilbert case, we find that the classical contribution to the  bending angle is universal, but unlike that case, universality is preserved also by the first quantum correction.    Finally we extend our analysis to include a deformation of the form $\Phi R^2$, where $\Phi$ is the dilaton, which arises in the low-energy effective action of the bosonic string in addition to the $R^3$ term, and compute its effect on the graviton bending. 
%We also comment on the potential relevance of quadratic corrections to gravity. 

	\vfill
	\hrulefill
	\newline
\vspace{-1cm}
$^{\alpha^\prime}$~\!\!{\tt\footnotesize\{a.brandhuber, g.travaglini\}@qmul.ac.uk}	
	
	\setcounter{page}{0}
	\thispagestyle{empty}
	\newpage

	%%%%%%%%%%%%%%%%%% TABLE OF CONTENTS %%%%%%%%%%%%%%%%%%%%%%%%%%%%%%%%%

	\setcounter{tocdepth}{4}
	\hrule height 0.75pt
	\tableofcontents
	\vspace{0.8cm}
	\hrule height 0.75pt
	\vspace{1cm}
	
	\setcounter{tocdepth}{2}

	\newpage
	%%%%%%%%%%%%%%%%%%%%%%%%%%%%%%%%%%%%%%%%%%%%%%%%%%%%%%%%%%%

\section{Introduction}

Modern on-shell methods \cite{Bern:1994zx,Bern:1994cg}  have proven extremely successful for the efficient  computation of scattering amplitudes  in gauge theory and gravity.  By working with on-shell quantities one performs computations which are at every stage gauge invariant, yielding  considerable conceptual and practical advantages. 

Recently, amplitude methods have been applied to the computation of post-Newtonian and post-Minkowskian corrections in General Relativity (GR).  Examples include the computation of the leading classical  \cite{Neill:2013wsa,Bjerrum-Bohr:2013bxa} and quantum  \cite{Bjerrum-Bohr:2013bxa} corrections  at $\cO(G_N^2)$ to the Newton potential, confirming the earlier result of \cite{BjerrumBohr:2002kt,Khriplovich:2004cx,Iwasaki:1971vb} based on Feynman diagrams, as well as the computation of the  particle bending angle \cite{Bjerrum-Bohr:2014zsa, Bjerrum-Bohr:2016hpa, Bai:2016ivl, Chi:2019owc} (for other recent related computations see 
\cite{Luna:2016due,Cachazo:2017jef,Bjerrum-Bohr:2018xdl,Plefka:2018dpa,Cheung:2018wkq,Kosower:2018adc,  Guevara:2018wpp,Bern:2019nnu,  Bautista:2019tdr, KoemansCollado:2019ggb}). This is clearly a timely endeavour as  LIGO necessitates computations in GR of unprecedented precision.  Feynman diagram  calculations
have been employed for many years to extract relevant quantities for astrophysical processes.
%the potential in the effective two-body problem. 
In this context, gravity is treated as an  effective field theory  \cite{Donoghue:1994dn}, 
making it  perfectly  sensible to compute quantum  corrections even if the theory is  non-renormalisable. 
An alternative, systematic effective field theory treatment was introduced in \cite{Goldberger:2004jt}, where the massive objects are treated as classical sources. 
The main focus for LIGO applications is to compute classical corrections, which, due to an interesting cancellation of $\hbar$ factors, are in fact obtained  through loop calculations \cite{Holstein:2004dn}.  
Notable  efforts  include the computations of the  Newton  potential  at  second \cite{Damour:1985mt,Gilmore:2008gq}, third \cite{Damour:2001bu, Blanchet:2003gy, Itoh:2003fy,Foffa:2011ub}, fourth \cite{Jaranowski:2012eb, Damour:2014jta,Damour:2015isa, Damour:2016abl, Bernard:2015njp, Bernard:2016wrg, Foffa:2012rn, Foffa:2016rgu} and fifth \cite{Foffa:2019hrb,Blumlein:2019zku} post-Newtonian order, following the landmark  computation at first post-Newtonian order \cite{Einstein:1938yz}. Note also the effective one-body approach of \cite{Buonanno:1998gg},  recently extended to incorporate the first and second post-Minkowskian corrections in \cite{Damour:2016gwp,Damour:2017zjx}, respectively.

% In this paper we want to prove string theory ;-)
In this paper we entertain the possibility of adding higher-derivative curvature terms to the Einstein-Hilbert (EH)  action that could arise either
from string theory or other ultraviolet  completions of gravity, and consider their effect on two quantities of relevance: the Newton potential, and the particle bending angle. 
Concretely, we will consider the action
\beq
\label{action}
S \ = \ -{2\over \kappa^2} 
\int\!d^4x \sqrt{ -g}  \Big[ R  \ + \ 
{\alpha^{\prime \, 2} \over 48} \, I_1 \, + \, {\alpha^{\prime \, 2} \over 24} \, G_3 
\Big] \ ,
\eeq
where $I_1 := {R^{\alpha \beta}}_{\mu \nu} {R^{\mu \nu}}_{\rho \sigma} {R^{\rho \sigma}}_{\alpha \beta}$, and 
\beq
\label{G3bis}
G_3 := I_1 - 2 {R^{\mu \nu \alpha}}_\beta {R^{\beta \gamma}}_{\nu \sigma} {R^\sigma}_{\mu \gamma \alpha}\, . 
\eeq
Here $\alpha^\prime$ has dimension length squared,  $\kappa^2 = 32 \pi G_N$, and  $G_N$ is   Newton's constant.
We now briefly discuss the two cubic terms we have added to the EH action. 

The first cubic coupling, the second term in the action \eqref{action},  has a very special feature: it is the only $R^3$-invariant that affects three- and four-graviton amplitudes \cite{vanNieuwenhuizen:1976vb,Broedel:2012rc}; in particular it produces three-graviton amplitudes with all-plus or all-minus helicities, in addition to the single-minus and single-plus tree amplitudes coming from the EH term. This term is also the two-loop counterterm for pure gravity, although in the following we use it as a tree-level deformation of the EH action. A number of  amplitudes in this theory were computed in \cite{Broedel:2012rc}, also in the light of KLT relations  \cite{Kawai:1985xq} and the BCJ double-copy construction \cite{Bern:2008qj}.

The second cubic coupling, the third term in  \eqref{action},  has been introduced to take into account the other possible contraction of three Riemann tensors 
${R^{\mu \nu \alpha}}_\beta {R^{\beta \gamma}}_{\nu \sigma} {R^\sigma}_{\mu \gamma \alpha}$, whose contribution to the 
Newton potential was recently computed  in \cite{Emond:2019crr}. 
 As it turns out, a more natural combination to consider is $G_3$ defined above in \eqref{G3bis}. There are several reasons for this: 
first, $G_3$  appears in the low-energy effective action of the  bosonic string (which we quote later in Section \ref{actionstring}),  
and it is a topological invariant in six dimensions.  Furthermore,  its three- and four-point graviton amplitudes vanish  \cite{vanNieuwenhuizen:1976vb,Broedel:2012rc}. 

Together, $I_1$ and $G_3$ are the only two independent dimension-six couplings up to field redefinitions as far as $S$-matrix elements are concerned   \cite{Metsaev:1986yb,BjerrumBohr:2003vy}. 
Note that we  have introduced the two couplings $I_1$ and $G_3$ in \eqref{action} with the particular coefficients arising from the bosonic string;  in practice we will analyse their effects separately, and one could give them arbitrary coefficients if one wishes to consider a more general effective action.  
Moreover,  in addition to the two independent cubic couplings discussed now, we will   also consider  a coupling of the form $\Phi R^2$,  which appears in the full low-energy effective action of the  bosonic string, where  $\Phi$ represents the dilaton.

A comment is in order here. In principle one can also   consider adding to the EH action  quadratic terms of the form $R^2$, $R^{\mu \nu}R_{\mu \nu}$ and $R^{\mu \nu\rho\sigma}R_{\mu \nu\rho\sigma}$ (or, instead of the latter, the Gau\ss-Bonnet combination   $R^{\mu \nu\rho\sigma}R_{\mu \nu\rho\sigma} - 4R^{\mu \nu}R_{\mu \nu} + R^2$, which is a total derivative in four dimensions). 
However, it turns out  \cite{Huber:2019ugz} that both  $R^2$ and  $R^{\mu \nu}R_{\mu \nu}$ terms can be removed from the action with a field redefinition,  which leaves scattering  amplitudes invariant as a consequence of  the $S$-matrix equivalence theorem \cite{Tseytlin:1986zz,Tseytlin:1986ti,Deser:1986xr,Metsaev:1986yb}.
Hence such terms can only give contact-term contributions which do not affect  the  Newton potential  \cite{Huber:2019ugz}.%
\footnote{Note that  in \cite{Julve:1978xn} quadratic corrections arising from the addition of terms of the form $R^2$ and $R^{\mu \nu}R_{\mu \nu}$ where treated exactly, and found to modify the spectrum of the EH theory by the addition of massive scalar and tensor modes, as well as tachyonic and ghost modes, depending on the coefficients of these couplings. 
The new propagators were then used in  
 \cite{Alvarez-Gaume:2015rwa} to compute corrections to the Newton potential at tree level. In the  approach pursued in this work we  treat  such terms as perturbations of the EH theory in an effective field theory expansion, as advocated in \cite{Donoghue:1994dn,Simon:1990ic,
      Simon:1990jn}, where the massive modes simply do not propagate.}

Coming back to the main thread of this paper, we will focus on the computation of the following  two quantities of interest: first, the leading classical and quantum corrections to the Newton potential between two massive scalars, and second, the bending angle of massless particles of spin 0, 1 and 2 in the background of a heavy scalar. We extract these quantities from two-to-two scattering amplitudes at one loop which, as is well known in the literature \cite{Holstein:2004dn}, contains both classical and quantum corrections. Compared to the EH case we observe a further power suppression in the potentials consistent with the higher-derivative nature of the operator. The result for the classical contribution to the bending angle is expected to be spin-independent due to the equivalence principle, while this is not expected at the quantum level. Indeed in Einstein gravity this has been confirmed by \cite{Bjerrum-Bohr:2014zsa, Bjerrum-Bohr:2016hpa, Bai:2016ivl, Chi:2019owc}. Surprisingly, we find that also the first quantum correction to the bending angle is independent of the scattered particle in the presence of an $R^3$ coupling.
For completeness of our presentation we will also  discuss  the  corrections to the Newton potential arising from \eqref{G3bis}, which are non-vanishing, in agreement with  \cite{Emond:2019crr}. In addition, we will   show that  the $G_3$ interaction does not contribute to the  bending of massless particles in the background of massive scalars. Finally,  
the only process that is affected by the addition of a $\Phi R^2$ coupling is the graviton bending, and we will also compute the modification induced by this term. 

Note that we use this action as a low-energy effective theory, as the processes under consideration involve small energies and momenta, and is valid even if $\sqrt{\alpha'} \gg \kappa \sim \ell_{\mathrm{pl}}$ as is the case in string theory. This possibility can enhance the effect of the $R^3$-corrections  significantly compared to the more standard choice $\sqrt{\alpha'} \sim \kappa$.
In the context of gravitational wave experiments we do not expect the corrections arising from $R^3$ terms to quantities such as the Newton potential to be accessible because of the large distance scales involved, and it would clearly be of great interest to find instances where they could play a role. We also note  \cite{Camanho:2014apa}, where a  detailed analysis of causality constraints on the modifications of three-graviton interactions in the regime of large $\alpha^\prime$ was carried out, and the consequences for possible ultraviolet  completions of the effective gravity theory were studied.

The rest of the paper is organised as follows. In the next section we compute the classical and quantum correction to Newton's potential to order $(\alpha^\prime G_N)^2$. Section~\ref{sec:paben} is devoted to the calculation of the bending angle for particles of spin 0, 1 and 2 scattered off a heavy scalar. As anticipated, to order $(\alpha^\prime G_N)^2$ we find that the classical and quantum corrections to the bending angle are independent of the spin of the scattered particles. The universality of the classical part is a consequence of the equivalence principle; that of the quantum part deserves further exploration. Also in that section we consider the new contribution to the graviton bending angle due to the inclusion of a coupling of the form $\Phi R^2$, which  arises in the bosonic string theory.  Section~\ref{sec:end} contains our  concluding remarks. 
We include in Appendix~\ref{appen}  the expressions of the integral functions and Fourier transforms used throughout the paper.

\section{$R^3$ corrections to the gravitational  potential}
\label{sec:potential}

In this section we compute the leading classical and quantum corrections to the Newton potential induced by  adding  $R^3$ couplings  to the EH action \eqref{action}. 

Following \cite{Neill:2013wsa,Bjerrum-Bohr:2013bxa} (see also earlier work in \cite{BjerrumBohr:2002kt}), the potential can efficiently be
obtained from the computation of  the scattering amplitude 
of two scalar particles with masses $m_1$ and $m_2$. 
In the case of our interest, namely corrections due to the $R^3$ term in \eqref{action}, it turns out that surprisingly the Born term is absent and the leading classical and quantum corrections arise at one loop. We will perform this calculation efficiently with well-established unitarity methods for amplitudes.
The same approach will be used in the next section to determine the bending of a massless scalar by taking one of the two masses to zero. 

In order to set the stage for the calculation we first discuss the kinematics of the $2 \to 2$ scattering process.
To align with the notation used in subsequent sections we will choose the particle momenta so that $p_1^2 = p_2^2 = m_1^2$, $p_3^2 = p_4^2 = m_2^2$. We choose to parametrise the external momenta in the centre-of-mass frame as follows:
\begin{align}
\begin{split}
\label{kinematics}
p_1^\mu & =  -(E_1,\vec{p}-\vec{q}/2) \, ,  \\
p_4^\mu & = - (E_4,-\vec{p}+\vec{q}/2) \, , \\
p_2^\mu & =  (E_2,\vec{p}+\vec{q}/2) \, ,  \\
p_3^\mu & =  (E_3,-\vec{p}-\vec{q}/2)\ .
\end{split}
\end{align}
Furthermore, since we are considering elastic scattering we have 
\begin{align}
\label{ener}
\begin{split}
E_1&=\!E_2\!=\!\sqrt{m_1^2 +\vec{p}^{\, \,2}+\vec{q}^{\, \,2}/4}\ , 
\\
E_3&=\!E_4\!=\!\sqrt{m_2^2 +\vec{p}^{\, \, 2}+\vec{q}^{\, \,2}/4}\ , 
\end{split}
\end{align}
 where  $\vec{p} \, \cdot \, \vec{q}=0$ due to momentum conservation. Notice that due to our all-outgoing convention for the external lines, the four-momenta $p_1$ and $p_4$, corresponding to the incoming particles, have an overall sign. 
 Furthermore, our Mandelstam variables are defined as:
\begin{align}
\label{mandel}
s:=(p_1+p_2)^2 = -\vec{q}^{\, \, 2}   , \ \  t:=(p_1+p_4)^2 = (E_1+E_4)^2  ,  
 \ \ u:=(p_1+p_3)^2  , 
 \end{align}
with $s+t+u = 2 ( m_1^2 + m_2^2)$.
In this notation, the spacelike momentum transfer squared is given by $s$, while the centre of mass
energy squared is given by $t$.

A comment is in order here. We will later be interested in computing the classical and one-loop quantum contributions to the potential%
\footnote{To be precise, by this we mean the $\hbar^0$ and $\hbar^1$ terms of the potential. Due to the presence of massive particles the power of $\hbar$ is not related to the number of loops \cite{Holstein:2004dn}.} 
arising from a (in this case leading) one-loop computation. This  is obtained from the appropriately normalised amplitude by means of a Fourier transform in $\vec{q}$ \cite{Iwasaki:1971vb}. Reinstating powers of $\hbar$, this Fourier transform involves a factor of $\exp (i  \vec{q}\cdot \vec{r} / \hbar)$. It is important to  be able to disentangle classical and quantum effects, and this  can be achieved efficiently by replacing  $\vec{q}= \hbar \vec{k}$ and then integrating over the wavevector, as 
carefully discussed in \cite{Kosower:2018adc}. This in turn implies that we can suppress the term $\vec{q}^{\, \, 2}$  in the expression of the energies in \eqref{ener}, which would produce $\cO (\hbar^2)$ corrections. Similarly, in the following we will suppress such corrections from expanding the Mandelstam variables $t$ or $u$.

Moving on to the unitarity-based calculation of the scattering process, we stress a crucial fact, namely that classical and quantum corrections to the potential are associated with terms in the amplitude that are non-analytic in the variable $s$ \cite{Donoghue:1994dn, Holstein:2004dn} and, hence, have discontinuities in $s$. Therefore, it will suffice to consider two-particle cuts in the $s$-channel, see {\it e.g.} \cite{Neill:2013wsa,Bjerrum-Bohr:2013bxa} where modern on-shell methods were applied for the first time to this kind of problem. Furthermore, we only need to perform the cuts in four dimensions as discrepancies with $D$-dimensional cuts at one loop are related to rational, and hence, analytic terms.

\begin{figure}[htbp]
\begin{center}
\begin{tikzpicture}[line width=1.5 pt,scale=.9]
	\draw[double, fermionnoarrow, line width = .75] (0:1.6) -- (-30:3);
	\draw[double, fermionnoarrow, line width = .75] (0:1.6) -- (30:3);
	\draw[double, fermionnoarrow, line width = .75] (180:1.6) -- (150:3);
	\draw[double, fermionnoarrow, line width = .75] (180:1.6) -- (210:3);
	\draw  [red, dashed, line width = 1.0] (90:1.5) -- (270:1.5);
	\draw[fill=white] (180:1.6) circle (.6);
	\draw[fill=white] (0:1.6) circle (.6);
	\draw[double, vector, line width = .75] (-1.08,.25) arc (120:60:2.16);
	\draw[->, line width = .5] (-.35,1.0) arc (100:80:2.16);
	\draw[->, line width = .5] (-.35,-1.0) arc (-100:-80:2.16);
	\draw[double, vector, line width = .75] (1.08,-.25) arc (-60:-120:2.16);
        \node at (179:1.6) {\large $R^3$};
        \node at (0:1.6) {EH};
        \node at (215:3.4) {$1^{\phi_{m_1}}$};
        \node at (145:3.4) {$2^{\phi_{m_1}}$};
         \node at (34:3.4) {$3^{\phi_{m_2}}$};
          \node at (-34:3.4) {$4^{\phi_{m_2}}$};
           \node at (120:1.15) {$\ell_1$};
          \node at (-120:1.15) {$\ell_2$};
          \node at (180:3.5) {$m_1$};
          \node at (0:3.5) {$m_2  \ \ \ \    +$};
          \node at (90:1.6) {$\pm \ \ \ \mp$};
          \node at (-90:1.6) {$\pm \ \ \ \mp$};
\end{tikzpicture}
\begin{tikzpicture}[line width=1.5 pt,scale=.9]
	\draw[double, fermionnoarrow, line width = .75] (0:1.6) -- (-30:3);
	\draw[double, fermionnoarrow, line width = .75] (0:1.6) -- (30:3);
	\draw[double, fermionnoarrow, line width = .75] (180:1.6) -- (150:3);
	\draw[double, fermionnoarrow, line width = .75] (180:1.6) -- (210:3);
	\draw  [red, dashed, line width = 1.0] (90:1.5) -- (270:1.5);
	\draw[fill=white] (180:1.6) circle (.6);
	\draw[fill=white] (0:1.6) circle (.6);
	\draw[double, vector, line width = .75] (-1.08,.25) arc (120:60:2.16);
	\draw[->, line width = .5] (-.35,1.0) arc (100:80:2.16);
	\draw[->, line width = .5] (-.35,-1.0) arc (-100:-80:2.16);
	\draw[double, vector, line width = .75] (1.08,-.25) arc (-60:-120:2.16);
        \node at (179:1.6) {EH};
        \node at (0:1.6) {\large $R^3$};
        \node at (215:3.4) {$1^{\phi_{m_1}}$};
        \node at (145:3.4) {$2^{\phi_{m_1}}$};
         \node at (34:3.4) {$3^{\phi_{m_2}}$};
          \node at (-34:3.4) {$4^{\phi_{m_2}}$};
           \node at (120:1.15) {$\ell_1$};
          \node at (-120:1.15) {$\ell_2$};
          \node at (180:3.5) {$m_1$};
          \node at (0:3.5) {$m_2$};
          \node at (90:1.6) {$\pm \ \ \ \mp$};
          \node at (-90:1.6) {$\pm \ \ \ \mp$};
\end{tikzpicture}
\caption{The two cut diagrams contributing to the leading $R^3$ correction to the gravitational scattering of two massive scalars. The two gravitons crossing the cut have both either positive or negative helicity and we have indicated this next to the dashed lines.}
\label{R3Newton-fig}
\end{center}
\end{figure}
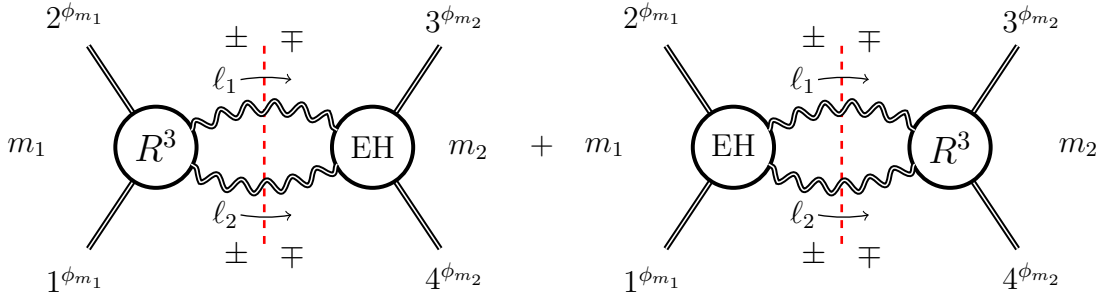

The relevant channel to consider is therefore that  associated with the momentum transfer in the scattering process. In this channel there are only two cut diagrams to consider, depicted in Figure~\ref{R3Newton-fig}. They are  related  by swapping the EH amplitude with the $R^3$ amplitude, which is equivalent to swapping $m_1$ and $m_2$ in the first diagram.

The cut calculation requires as input two types of two-scalar/two-graviton tree amplitudes.
While the corresponding tree amplitudes in EH gravity with a minimally coupled scalar are well known, we need to derive the expression for the amplitudes due to the $R^3$ correction. Note that this interaction forces the two internal gravitons to have equal helicities, since the $R^3$ term can only produce three-graviton amplitudes with all helicities equal.

%There is just one Feynman diagram contributing to this amplitude, see Figure ***. 

The well-known EH amplitude for the scattering of two scalars with mass $m_1$  and  two gravitons 
is given by  \cite{Bern:1998sv}
\beq
A( 1^{\phi_{m_1}}, 2^{\phi_{m_1}}, \ell_1^{--}, \ell_2^{--} ) \ = \ - \left({\kappa\over 2}\right)^2 \, m_1^4 {\langle \ell_1 \, \ell_2\rangle^2 \over [ \ell_1\, \ell_2]^2}\Big[ {i \over (\ell_1 + p_1)^2 - m_1^2}\, + \, {i \over (\ell_1 + p_2)^2 - m_1^2}\Big]\ . 
\eeq
The amplitude with two scalars of mass $m_2$ and two gravitons produced by one insertion of $R^3$ can easily be computed, with the result
\beq
\label{2s2grav}
A_{R^3}(-\ell_1^{++}, -\ell_2^{++}, 3^{\phi_{m_2}}, 4^{\phi_{m_2}}) \ = \ \Big({  \kappa\over 2} \Big)^2 \Big({  \alpha^\prime \over 4} \Big)^2 {4 i\over  s_{12}} [\ell_1\, \ell_2]^4 \, (\ell_1\cdot p_3) (\ell_2\cdot p_3)
\,  .
\eeq
In order to arrive at  \eqref{2s2grav} we had to evaluate a single Feynman diagram, 
and  we used the expression of the three-point vertex with two scalars of mass $m$ and momenta $p_1$ and $p_2$  and one off-shell graviton%
 \footnote{See for instance  \cite{Holstein:2008sx}.}
%\beq
%\label{js}
%V^{\mu \nu}_{\phi_m \phi_m h} (p_1, p_2)  \ = \ i\left({  \kappa\over 2} \right)\Big[ -\eta^{\mu \nu} (p_1\cdot p_2 + m^2) + %p_1^\mu p_2^\nu + p_2^\nu p_1^\mu\Big]
%\ , 
%\eeq
\begin{align}
\label{js}
V^{\mu \nu}_{\phi_m \phi_m h} (p_1, p_2) \ =
\raisebox{-.48\height}{
\begin{tikzpicture}[line width=1. pt, scale=.5]
	\draw[double, vector, line width = .75] (0:0) -- (0:2);
	\draw[double,fermionnoarrow,  line width = .75] (0:0) -- (120:2);
	\draw[double,fermionnoarrow,  line width = .75] (0:0) -- (240:2);
        \node at (-90:2) {$1^{\phi_m}$};
        \node at (90:2) {$2^{\phi_m}$};
         \node at (0:2.6) {$\mu\nu$};
\end{tikzpicture}
}
= \ i\left({  \kappa\over 2} \right)\Big[ -\eta^{\mu \nu} (p_1\cdot p_2 + m^2) + p_1^\mu p_2^\nu + p_2^\nu p_1^\mu\Big]
\ , 
\end{align}
along with the three-point {\it current} $X^{\mu \nu}_{R^3} (1^{++}, 2^{++})$ with two on-shell, positive helicity gravitons and one off-shell graviton derived from the $R^3$ coupling, which is found to be
%\beq
%\label{corrente}
%X^{\mu \nu}_{R^3} (1^{++}, 2^{++})\ = \ {i\over 4} [1\, 2]^4 \big( \langle 1 | \mu | 2] \langle 2 | \nu | 1] \ + \ \mu \leftrightarrow \nu \big)
%\ .
%\eeq
\begin{align}
\label{corrente}
X^{\mu\nu}_{R^3}(1^{++},2^{++}) \ =
\raisebox{-.48\height}{
\begin{tikzpicture}[line width=1.5 pt,scale=.5]
	\draw[double, vector, line width = .75] (0:0) -- (0:2);
	\draw[double, vector, line width = .75] (0:0) -- (120:2);
	\draw[double, vector, line width = .75] (0:0) -- (240:2);
        \node at (60:1.2) {$R^3$};
        \node at (-90:2) {$1^{++}$};
        \node at (90:2) {$2^{++}$};
         \node at (0:2.6) {$\mu\nu$};
\end{tikzpicture}
}
= \ {i\over 4} \Big({  \kappa\over 2} \Big) \Big({  \alpha^\prime \over 4} \Big)^2 [1\, 2]^4 \big( \langle 1 | \mu | 2] \langle 2 | \nu | 1] \ + \ \mu \leftrightarrow \nu \big)\, .
\end{align}
%Note that  we have chosen to normalise \eqref{corrente}  in such a way that the three positive-helicity graviton amplitude is 
Note that this gives the well-known three positive-helicity graviton amplitude if we contract the free indices with the appropriate polarisation tensor, 
\beq
A_{R^3}( 1^{++}, 2^{++}, 3^{++} ) \ = \ -i \Big({  \kappa\over 2} \Big) \Big({  \alpha^\prime \over 4} \Big)^2 ([12][23][31])^2
\ . 
\eeq
%Later on we will reinstate the correct normalisations and $\kappa$ and $\alpha^\prime$ dependence, which amounts to
% the following: 
%\begin{itemize}
%\item[{\bf 1.}]~The vertex in \eqref{js} is multiplied by $\kappa/2$;
%\item[{\bf 2.}]~
%multiply the off-shell current \eqref{corrente}  by a factor of $(\alpha^\prime/4)^2 (\kappa/2)$ arising from the action \eqref{action}.
% and the expansion of the metric as $g = \eta + \kappa h$. 
%\end{itemize}

%
The two four-point amplitudes quoted above can now be combined to form the cut integrand  in the
$s$-channel.
Note that in our conventions all external particle momenta $p_i$ are considered as outgoing.
From the left-hand side of Figure \ref{R3Newton-fig} we get
\beq
\label{tcut}
\left. \mathcal{I}_{\phi_{m_1},\phi_{m_2}}^{(1),{\mathrm{LHS}}}\right|_{s\text{-cut}} \ = \  \mathcal{D} \, 4 \, m_1^4 \, s (\ell_1\cdot p_3) (\ell_2 \cdot p_3) 
\Big[ {1\over (\ell_1 + p_1)^2 - m_1^2} \, + \, {1\over (\ell_1 + p_2)^2 - m_1^2}\Big]\
\ , 
\eeq
where we have taken into account that the factor of two due to the sum over internal helicity configurations is exactly cancelled by a factor one half coming from the fact that two identical (same-helicity) particles are crossing the cut.
Furthermore, we
have introduced the universal combination of couplings
\beq
\mathcal{D} = \Big({  \kappa\over 2} \Big)^4 \Big({  \alpha^\prime \over 4} \Big)^2 \ ,
\eeq 
and we have suppressed the ubiquitous two-particle phase space measure.
The second cut diagram (right-hand side of Figure~\ref{R3Newton-fig}) is obtained from the first by swapping $m_1$ and $m_2$. 

Lifting \eqref{tcut} off the cut, {\it i.e.}~taking $\ell_{1,2}$ off-shell and replacing the two cut propagators by $(i/ \ell_1^2) (i / \ell_2^2)$,
we obtain a one-loop integral with a rather complicated numerator. 
The reduction to a linear combination of scalar Feynman integrals can be performed efficiently using {\tt LiteRed} \cite{Lee:2012cn, Lee:2013mka}.
In Appendix \ref{appen} we have given all integrals that are relevant for the computation of the potential, namely those with discontinuities in the $s$-channel, and in the expression of the amplitudes presented below we will only include such integrals.

%where $p_1\rightarrow (-m_1, \vec{q})$, $p_2\rightarrow (m_1, \vec{q})$ while $p_3\rightarrow (-m_2, -\vec{q})$, and $p_4\rightarrow (m_2, -\vec{q})$, with $|\vec{q}| << m_{1,2}$. Setting $q=p_1+p_2\to (0, \vec{q})$  
%we then work in the limit where 
%\beq
%s_{12} \rightarrow - {\vec{q}}^{\, \, 2}\, , \qquad s_{23} \to (m_1+ m_2)^2
%\ . 
%\eeq
From the first  diagram in Figure~\ref{R3Newton-fig} we obtain
\beq
A_{\phi_{m_1}, \phi_{m_2}}^{(1),{\mathrm{LHS}}} = c_3  (m_1, m_2) I_3 (s; m_1) \, + \,  c_2 (m_1, m_2)  I_2(s)  \ ,
\eeq
where we have  suppressed for the moment the overall factor $\mathcal{D}$.
The full Lorentz-invariant expressions of $c_2$ and $c_3$ are: 
\begin{align}
\label{fff}
\small
\begin{split}
c_3(m_1,m_2) =  \frac{2 s^2 m_1^4}{(4 m_1^2-s)^2} & 
\left[ 2 m_1^2 \left(m_1^4-2 m_1^2 \left(m_2^2+t\right)+\left(m_2^2-t\right)^2\right) \right. \\
 & \left. +s \left(-3 m_1^4+2 m_1^2 m_2^2+\left(m_2^2-t\right)^2\right) + s^2 \left(m_1^2-m_2^2+t\right) \right] \\
c_2(m_1,m_2) = \frac{s^2 m_1^4}{(4 m_1^2-s)^2} & 
\left[6 m_1^4+4 m_1^2 \left(m_2^2-3 t\right)+6 \left(m_2^2-t\right)^2 -2 s \left(2 \left(m_1^2+m_2^2\right)-3 t\right) +s^2 \right]. 
\end{split}
\end{align}
As discussed after \eqref{mandel},  we only need to keep the leading-order term in $s= - |\vec{q}\, |^{2}$ of \eqref{fff}. This is all what is needed in order to extract  the full post-Minkowskian   (classical plus one-loop quantum) potential. The resulting expressions are:
\begin{align}
\label{fffbis}
\small
\begin{split}
\tilde{c}_3(m_1,m_2) &=  \frac{ (m_1s)^2}{4} 
\left[ (t - m_1^2 - m_2^2)^2 - 4m_1^2 m_2^2 \right]\, , \\
%
%m_1^4-2 m_1^2 \left(m_2^2+t\right)+\left(m_2^2-t\right)^2  \right] \\
\tilde{c}_2(m_1,m_2)&= \frac{s^2}{8} 
\left[3  (t - m_1^2 - m_2^2)^2 -4 m_1^2 m_2^2 
%******
%6 m_1^4+4 m_1^2 \left(m_2^2-3 t\right)+6 \left(m_2^2-t\right)^2  
\right]. 
\end{split}
\end{align}
For convenience we also quote the result for the post-Newtonian expansion, which requires further expanding  for $|\vec{p}\, | \ll m_{1,2}$. 
In this non-relativistic limit, we have 
% t:=(p_1+p_4)^2 = (E_1+E_4)^2 \simeq (m_1+m_2)^2\left(1+ \frac{\vec{p}^{\, \,2} + 
%\vec{q}^{\, \,2}/4
%}{m_1 m_2} \right) \ ,  \\
% ************
%These are
%\begin{align} 
%\begin{split}
%c_3  (m_1, m_2)  & \simeq (m_1 s)^2 \Big[2 (m_1+m_2)^2 \vec{p}^{\, \, 2} -s(m_1^2-m_2^2)  \Big]\, , \\
%c_2 (m_1, m_2) & \simeq s^2 \left[2 m_1^2 m_2^2+3(m_1+m_2)^2 \vec{p}^{\, \, 2}- \frac{s}{2}(m_1^2-m_2^2)\right]
%\ , 
%\end{split}
%\end{align}
\begin{align} 
\begin{split}
\tilde{c}_3  (m_1, m_2)  & \simeq (m_1 s)^2  (m_1+m_2)^2 \vec{p}^{\, \, 2} \, , \\
\tilde{c}_2 (m_1, m_2) & \simeq s^2 \left[m_1^2 m_2^2+\frac{3}{2}(m_1+m_2)^2 \vec{p}^{\, \, 2} \right]
\ .
\end{split}
\end{align}
Curiously, in the static limit $\vec{p}^{\, \, 2} \to 0$ the leading term of $c_2$ is $\cO (s^2)$, while $c_3$ is of order $\cO (s^3)$ and hence further suppressed.
The expressions for the bubble integral $I_2(s)$ and the massive triangle integral $I_3(s; m)$ are given  in \eqref{integrals}.

The classical contributions to the potential are identified with the non-analytic $1/\sqrt{-s}$ contributions, arising uniquely from the  $I_3(s; m_{1,2})$ integral:
\begin{align}
\begin{split}
\label{cre}
A_{\phi_{m_1}, \phi_{m_2}}^{(1),{\mathrm{cl}}} & = -\frac{i}{32 \sqrt{-s}}\left( {\tilde{c}_3  (m_1, m_2) \over m_1} + {\tilde{c}_3  (m_2, m_1) \over m_2}\right)  \\ 
& = \ -\frac{i s^2}{32 \sqrt{-s}} \frac{m_1 + m_2}{4} \left[  (t-m_1^2-m_2^2)^2 - 4 m_1^2 m_2^2 \right]  \ ,    \\
& \simeq \ -\frac{i s^2}{32 \sqrt{-s}} (m_1+m_2)^3 \vec{p}^{\, \, 2}  \ ,
\end{split}
\end{align}
where the middle line represents the full relativistic classical contribution, while the last line gives the small velocity approximation.%
\footnote{For the rest of this section we   denote the non-relativistic limit of the full relativistic expression by  $\simeq$.}

On the other hand the finite $\log(-s)$  terms from $I_2$ and $I_3$ are genuine quantum corrections:
\begin{align}
\label{qre}
A_{\phi_{m_1}, \phi_{m_2}}^{(1),{\mathrm{qu}}} &= 
- {i \over 16 \pi^2} s^2 \log (-s) \left[  (t-m_1^2-m_2^2)^2 - 2 m_1^2 m_2^2 \right] \nonumber \\ 
& \simeq - {i \over 8 \pi^2} s^2 \log (-s) \Big[
%
 %-i \frac{s^2 \log(-s)}{8 \pi^2}
 %
  m_1^2 m_2^2\, 
 %+ \, 4 \, s \, (m_1+m_2)^2 
+ 2 (m_1 + m_2)^2  \vec{p}^{\, \, 2}
 \Big]\ .
\end{align}
Finally, we  extract the gravitational potential  from the  three-dimensional  Fourier transform in $\vec{q}$ of the amplitude \cite{Iwasaki:1971vb}, 
\begin{align}
\label{potential}
V(\vec{r},\vec{p}) = i \int\!\frac{d^3q}{(2 \pi)^3} \ e^{i \vec{q} \cdot \vec{r}} \ \frac{A(\vec{q},\vec{p})}{4\,  E_1 \, E_4} \, ,
\end{align}
with $\vec{q}$ and $\vec{p}$ related to the Mandelstam variables as described earlier in \eqref{mandel}. 
We then get
\begin{align}
\label{potentialbis}
V(\vec{r},\vec{p}) &:= V_{\rm cl}(\vec{r},\vec{p}) \, + \, \hbar \, V_{\rm qu}(\vec{r},\vec{p}) \ = \ \int\!\frac{d^3q}{(2 \pi)^3} \ e^{i \vec{q} \cdot \vec{r}} \ \big( v_{\rm cl}  +\hbar v_{\rm qu}  \big) \, ,
\end{align}
with 
\begin{align}
\begin{split}
v_{\rm cl} & = \ \frac{s^2}{\sqrt{-s}} \frac{m_1 + m_2}{512 E_1 E_4} \left[  (t-m_1^2-m_2^2)^2 - 4 m_1^2 m_2^2 \right]   
 \simeq  \ \frac{s^2}{\sqrt{-s}} {(m_1 + m_2)^3 \vec{p}^{\, \, 2}  \over 128 m_1 m_2}  \, , 
\\ 
v_{\rm qu} & =  {1 \over 64 \pi^2} s^2 \log (-s) {\left[  (t-m_1^2-m_2^2)^2 - 2 m_1^2 m_2^2 \right] \over E_1 E_4}  \\ 
& \simeq \ {1\over 64 \pi^2} s^2 \log (-s ) \Big[ 2 m_1 m_2  + 
\vec{p}^{\, \, 2} \Big( 8 + 3  {m_1^2+m_2^2 \over m_1 m_2} 
\Big)
 \Big]
\ . 
\end{split}
\end{align}
Finally, we  reinstate the overall factor 
$\mathcal{D}=(\alpha^\prime/4)^2 (\kappa/2)^4$, introduce Newton's constant
$G_{N} :=\kappa^2 / (32 \pi)$, and perform the Fourier transforms using \eqref{a3} and \eqref{a4}. This gives our result for the leading classical and quantum corrections to Newton's potential arising from the addition of the $I_1$ coupling in \eqref{action} to Einstein's gravity: 
\begin{align}
\begin{split}
\label{clpot}
%\small
V_{\rm cl}(\vec{r},\vec{p}) & =
{( \alpha^\prime G_N)^2 \over r^6} \frac{3(m_1 + m_2)}{32 E_1 E_4} \left[  (t-m_1^2-m_2^2)^2 - 4 m_1^2 m_2^2 \right] \\
& \simeq  {( \alpha^\prime G_N)^2 \over r^6} \ \left[ 
{3\over 8}{(m_1 + m_2)^3\over m_1 m_2}  \, \vec{p}^{\, \, 2} \right]\ , 
\end{split}
\end{align}
and 
\begin{align}
\begin{split}
\label{qupot}
%\small
V_{\rm qu}(\vec{r},\vec{p}) &= {( \alpha^\prime G_N)^2 \over r^7} \left\{ - {15\over 4 \pi} {\left[  (t-m_1^2-m_2^2)^2 - 2 m_1^2 m_2^2 \right] \over E_1 E_4} \right\}  \\
& \simeq {( \alpha^\prime G_N)^2 \over r^7} \left\{ - {15\over 4 \pi} \Big[ 2 m_1 m_2 + \vec{p}^{\, \, 2} \Big( 8 + 3  {m_1^2+m_2^2 \over m_1 m_2} 
\Big)\Big] 
\right\}  . 
\end{split}
\end{align}

\noindent
%\subsection{Second $R^3$ structure}
While we discussed so far the effects of the interaction
$I_1 = {R^{\alpha \beta}}_{\mu \nu} {R^{\mu \nu}}_{\rho \sigma} {R^{\rho \sigma}}_{\alpha \beta}$, there
exists a second independent contraction ${R^{\mu \nu \alpha}}_\beta {R^{\beta \gamma}}_{\nu \sigma} {R^\sigma}_{\mu \gamma \alpha}$. Corrections to the Newton potential due to this  interaction were recently studied in \cite{Emond:2019crr}. These two structures combine naturally into   
\beq
\label{G3}
G_3 := I_1 - 2 {R^{\mu \nu \alpha}}_\beta {R^{\beta \gamma}}_{\nu \sigma} {R^\sigma}_{\mu \gamma \alpha}\ ,
\eeq
which appears in the low-energy effective action of the bosonic string (quoted later on in this paper  in \eqref{action-long}). It   is a topological invariant in six dimensions and its three- and four-point graviton amplitudes vanish  \cite{vanNieuwenhuizen:1976vb,Broedel:2012rc}.  
For completeness we will now  present   a short discussion of the  corrections to the Newton potential in the presence of the $G_3$ interaction.

The steps in the derivation of the potential are identical to the ones detailed above, but an important new ingredient is the two graviton/two scalar amplitude induced by the $G_3$-interaction with unit coefficient:
\begin{align}
\label{ag3}
A_{G_3}(1^{++},2^{++},3^{\phi_m},4^{\phi_m}) = -i { 3! \over 4} \Big({  \kappa\over 2} \Big)^4 
%\Big({  \alpha^\prime \over 4} \Big)^2
 [12]^4 (s+2 m^2) \ . 
\end{align}
Importantly, this expression contains a contribution  proportional to $m^2$ that leads to a qualitatively new term in the potential, while the term proportional to $s$ only gives a higher order in $\hbar$ correction which we will drop. Note also the absence of a collinear singularity in \eqref{ag3}; indeed the three-point graviton amplitudes generated by $G_3$ vanish.

Feeding the amplitude in \eqref{ag3} in the cut computation as done earlier leads to the amplitude for the scattering of two massive scalars with masses $m_{1,2}$: 
\begin{align}
3! \left({\kappa\over 2}\right)^6 (m_1 m_2 s)^2\left[ m_1^2  I_3(s,m_1) + m_1 \leftrightarrow m_2 \right] \ ,
\end{align}
where as usual we only kept the leading term in $s$.

The coupling $G_3$ appears in the low-energy bosonic string  effective action quoted later in \eqref{action-long} in the form  
 $\cL^\prime = (-2/\kappa^2) \alpha^{\prime \, 2} (G_3/24)$. For this particular interaction term,  
 going through the standard procedures 
one arrives at the  following  corrections to the potential: 
\begin{align}
V_{\cL^\prime} \ = \ -\frac{3}{4} (\alpha^\prime G_N)^2  {(m_1 m_2)^2 \over E_1 E_4}\, 
\left[(m_1+m_2) {1\over r^6}\, -\, \hbar \,  {10 \over \pi r^7}   \right] \ , 
\end{align}
where as usual $G_{N}$ is Newton's constant, 
and as before we have only written the classical
contribution and the first quantum correction.
Note one interesting difference between the classical correction arising from  $I_1$ and $G_3$, namely that  the latter does not vanish in the static limit $|\vec{p}\, | \to 0$.%
\footnote{In the non-relativistic limit, one can set 
$(m_1 m_2)^2 / (E_1 E_4) \to 
m_1 m_2- |\vec{p} \, |^2 \left(m_1^2+m_2^2\right)/ ( 2 m_1
   m_2)$.}

Finally, we anticipate that there is no contribution to the 
bending of massless particles from massive scalars in the presence of the $G_3$ coupling, 
as discussed in the next sections.

\section{Particle bending angle} 
\label{sec:paben}

In this section we  compute the effect of the $R^3$ term to the bending of massless particles of spin 0, 1 and 2  in the presence of a heavy scalar particle of mass $m$ using similar methods as in the previous section. We will compute the relevant  scattering amplitudes of massless scalars, photons and gravitons off a massive scalar in Sections \ref{sec:scalar}, \ref{sec:photon} and \ref{sec:graviton},  respectively, and then compute the bending angle in Section \ref{sec:bending}. 
Since we only consider elastic scattering, the helicity of the bent particle does not change in the process. Due to our convention that all particles have outgoing momenta, helicity conservation requires that the incoming massless particle has opposite helicity compared to the outgoing one.

Before starting it  is useful  to revisit the kinematics introduced in \eqref{kinematics} in the situation where  $m_1 \to m$ and $m_2 \to 0$.  In  this case we have 
\begin{align}
\begin{split}
\label{enbend}
E_1=E_2 =\sqrt{m^2+\vec{p}^{\, \, 2}+\vec{q}^{\, \, 2}/4}\, , 
\\
E_3=E_4=\sqrt{\vec{p}^{\, 2}+\vec{q}^{\, \, 2}/4}\,  := \, \omega
\ .
\end{split}
\end{align}
We then find that $s = -\vec{q}^{\, \, 2}$, as before, while  $t = (E_1+E_4)^2 \simeq m (m+2 \omega)$, and $u=2 m^2 - s -t$.
In order to extract the particle bending we work in a limit where 
\beq
\label{bendlim}
-s=\vec{q}^{\, \, 2} \ll \omega^2 \ll m^2
\ , 
\eeq
which also  implies
$ut - m^4 \simeq - (2 m \omega)^2$. 

\subsection{Scalar bending }
\label{sec:scalar}

The result for the bending of a massless scalar particle when it passes near a heavy scalar of mass $m$ can  be extracted from considering the right-hand side diagram in Figure~\ref{R3Newton-fig}, setting $m_2\to 0$ and renaming $m_1\to m$. The left-hand side diagram simply vanishes in this limit.
Doing so, and working in the limit \eqref{bendlim}, we arrive at the simple result 
\beq
\label{scalarbending}
A^{(1)}_\phi   =
%\left({\kappa\over 2}\right)^3 \, 
\mathcal{D} N_\phi 
\left[  (m^2 s\, \omega )^2  I_3 (s;m) 
\ + \frac{3}{2} (m s\, \omega )^2    I_2 (s) \right] \ , 
\eeq
where $N_\phi = 1$ is introduced only in order to then compare with the photon and graviton bending results in \eqref{photonbending} and \eqref{gravitonbending}. The expressions for the integral functions can  be found in Appendix~\ref{appen}. 

It is interesting to compare our result to the corresponding result for scalar bending in Einstein gravity, Eq.~(10) of \cite{Bjerrum-Bohr:2014zsa}. Our result contains two more powers of $s$, as expected from working with an $R^3$ interaction, which contains four more derivatives with respect to the EH action. As we will see later in \eqref{photonbending} and \eqref{gravitonbending}, we will arrive at a  result for the particle bending which is the same for scalars, photons and gravitons up to {\it and including} the first quantum correction. This universality of the quantum correction is  unexpected -- it is not a feature of Einstein gravity \cite{Bjerrum-Bohr:2014zsa, Bjerrum-Bohr:2016hpa, Bai:2016ivl, Chi:2019owc} -- and deserves further investigation. 

A final comment is in order. Due to the mass dependence in \eqref{ag3}, there are no classical and $\cO ( \hbar )$ corrections to the bending of {\it massless} scalars   due to the  $G_3$  coupling  in \eqref{G3} --  this is clear from \eqref{ag3}, where the $m^2$ term in the parenthesis vanishes while the second can be discarded because it induces  corrections of   ${\cal O} (\hbar^2)$.

\subsection{Photon  bending}
\label{sec:photon}
\begin{figure}[htbp]
\begin{center}
\begin{tikzpicture}[line width=1.5 pt, scale=.9]
	\draw[vector] (0:1.6) -- (-30:3);
	\draw[vector] (0:1.6) -- (30:3);
	\draw[double, fermionnoarrow, line width = .75] (180:1.6) -- (150:3);
	\draw[double, fermionnoarrow, line width = .75] (180:1.6) -- (210:3);
	\draw  [red, dashed, line width = 1.0] (90:1.5) -- (270:1.5);
	\draw[fill=white] (180:1.6) circle (.6);
	\draw[fill=white] (0:1.6) circle (.6);
	\draw[double, vector, line width = .75] (-1.08,.25) arc (120:60:2.16);
	\draw[->, line width = .5] (-.35,1.0) arc (100:80:2.16);
	\draw[->, line width = .5] (-.35,-1.0) arc (-100:-80:2.16);
	\draw[double, vector, line width = .75] (1.08,-.25) arc (-60:-120:2.16);
        \node at (179:1.6) {EH};
        \node at (0:1.6) {\large $R^3$};
        \node at (215:3.4) {$1^{\phi_{m}}$};
        \node at (145:3.4) {$2^{\phi_{m}}$};
%         \node at (34:3.4) {$p_3$};
%          \node at (-34:3.4) {$p_4$};
                   \node at (34:3.4) {$3^+$};
          \node at (-34:3.4) {$4^-$};
           \node at (120:1.15) {$\ell_1$};
          \node at (-120:1.15) {$\ell_2$};
          \node at (90:1.6) {$\pm \ \ \ \mp$};
          \node at (-90:1.6) {$\pm \ \ \ \mp$};
\end{tikzpicture}
\caption{The cut diagram contributing to the leading $R^3$ correction to gravitational scattering of a photon (wavy lines) off a massive scalar (double lines).}
\label{photon-bending-fig}
\end{center}
\end{figure}
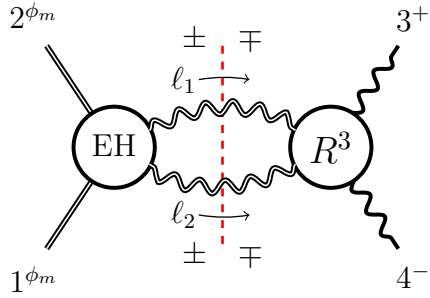

The  cut diagram to compute in this case is shown in Figure~\ref{photon-bending-fig}. The amplitudes entering the cut are 
\beq
\label{first}
A( 1^{\phi_m}, 2^{\phi_m}, \ell_1^{--}, \ell_2^{--} ) \ = \ - \left({\kappa\over 2}\right)^2 \, m^4 {\langle \ell_1 \, \ell_2\rangle^2 \over [ \ell_1\, \ell_2]^2}\Big[ {i \over (\ell_1 + p_1)^2 - m^2}\, + \, {i \over (\ell_1 + p_2)^2 - m^2}\Big]\ , 
\eeq
while for the two-photon/two-graviton amplitude we have 
\beq
\label{prima} 
A_{R^3} ( -\ell_2^{++}, -\ell_1^{++}, 3^{+}, 4^{-} ) \ = \     - i \Big({ \kappa\over 2} \Big)^2
\Big( {\alpha^\prime \over 4}\Big)^2   { [\ell_1 \ell_2]^4 \over s_{12}} \, \langle 4 | \ell_1 | 3]^2
\ . 
\eeq
The latter amplitude can be derived by using the expression of the minimal two-photon/one graviton coupling (see for instance Section (4.4) \cite{Schuster}), which for the required helicities simplifies to
%\beq
%V_{\mu \nu} (1^+, 2^- ) \ = \  -  {i\over 2} \left(  {\kappa\over 2}\right)  \, \langle 2 | \mu | 1] \, \langle 2 | \nu | 1] 
%\ . 
%\eeq
\begin{align}
V^{\mu\nu}(1^{+},2^{-}) \ =
\raisebox{-.48\height}{
\begin{tikzpicture}[line width=1. pt,scale=.5]
	\draw[double, vector, line width = .75] (0:0) -- (0:2);
	\draw[vector] (0:0) -- (120:2);
	\draw[vector] (0:0) -- (240:2);
        \node at (-90:2) {$1^{+}$};
        \node at (90:2) {$2^{-}$};
         \node at (0:2.6) {$\mu\nu$};
\end{tikzpicture}
}
= \ {i\over 2} \left(  {\kappa\over 2}\right)  \, \langle 2 | \mu | 1] \, \langle 2 | \nu | 1] \ .
\end{align}
Contracting this with the already derived current \eqref{corrente} with two same-helicity gravitons and an additional off-shell 
graviton via   
the standard de Donder propagator      leads to \eqref{prima}. 

Using \eqref{first} and \eqref{prima} we arrive at the following expression for the cut integrand
\beq
\label{conLC}
\left. \mathcal{I}^{(1)}_{\gamma} \right|_{s\textrm{-cut}} \ = \    \mathcal{D} s \, m^4  \langle 4| \ell_1 | 3]^2 \Big[ {1\over (\ell_1 + p_1)^2 - m^2} + {1\over (\ell_1 + p_2)^2 - m^2} \Big]
\ ,
\eeq
corresponding to the cut diagram in Figure \ref{photon-bending-fig}.  
Reductions can be performed using the identity 
\beq
\label{identity}
 \langle 4 | \ell_1 | 3] =- {{\rm Tr}_{+} ( 4 \ell_1 3 1) \over  \langle 3| 2| 4]}\ , 
 \eeq
%\beq
%\langle 4 | \ell_1 | 3]^4 \ = \ \left[{2\,  \langle 4 | 1 | 3] \over ( s_{13} s_{23} - m^4)}\right]^4  
%\Big[  L - i \eps ( 4 \ell_1 3 1 ) \Big]^4
%\ , 
%\eeq 
so that the integrand taken off the cut becomes
\beq
\label{senzaLC}
\mathcal{I}^{(1)}_{\gamma}  = -  \mathcal{D} {4\, m^4 s_{12} \over \langle 3 | 2 | 4]^2 } \Big[ L^2 + E^2\Big]
\Big[ {1\over (\ell_1 + p_1)^2 - m^2} + {1\over (\ell_1 + p_2)^2 - m^2} \Big] \frac{1}{\ell_1^2} \frac{1}{\ell_2^2}
\ , 
\eeq
where 
\begin{align}
\begin{split}
L & := (p_1p_3) (p_4 \ell_1) - (p_3p_4) (p_1 \ell_1) + (p_4p_1) (p_3 \ell_1) \ , 
\\
E^2 &:= - \big[\eps ( p_4 \ell_1 p_3 p_1 ) \big]^2 = {\rm det} M \ , 
\end{split}
\end{align}
and $M$ is the matrix whose entries are the scalar products of the momenta in the $\eps$ symbol. In going from \eqref{conLC} to \eqref{senzaLC} we have also dropped terms linear in the Levi-Civita symbol, which vanish upon integration. 
After performing the tensor reduction, keeping only terms with an $s$-channel discontinuity and dominant in the limit
\eqref{bendlim}, we arrive at the simple  result  
\beq
\label{photonbending}
A^{(1)}_\gamma = 
-\mathcal{D} \, N_\gamma 
\left[ (m^2 s\, \omega )^2  I_3 (s;m) 
\ + \frac{3}{2}(m s\, \omega )^2    I_2 (s) \right] \ , 
\eeq
where $N_\gamma := \big[ (2 m \omega) / \langle 3| 2| 4]\big]^2$.
Note that  in the low-energy limit \eqref{bendlim} we have  $ \langle 3| 2| 4 ] \to i \langle 3| 2| 3 ] = i (t-m^2) = i (2 m \omega)$ hence 
\begin{equation}
\label{ngamma}
N_\gamma\to -1\ ,
\end{equation}
in this limit.  
%$ |\langle 3| 2| 4]|^2 = - ut+ m^4 \to (2 m \omega)^2$ in the low-energy limit \eqref{bendlim}. As observed in 
%\cite{Bjerrum-Bohr:2014zsa},  
%in this limit $N_\gamma$ is a  phase that does  not affect the potential and bending to  be derived in Section 
%\ref{sec:bending}. 

Comparing our result to that of light bending in Einstein gravity obtained in \cite{Bjerrum-Bohr:2014zsa}, we see that our result is suppressed by two powers of $s$ compared to \cite{Bjerrum-Bohr:2014zsa}, as expected from working with an $R^3$ interaction. 
Furthermore, we see that the term in square brackets in \eqref{photonbending} is identical to the corresponding term in \eqref{scalarbending}. This is true for the classical term (the massive triangle), in line with the equivalence principle, but also for the first quantum correction (the bubble contribution). 

Finally, in the presence of a $G_3$ interaction the tree-level amplitude on the right-hand side of Figure~\ref{photon-bending-fig} vanishes, that is  $A_{G_3} ( -\ell_2^{++}, -\ell_1^{++}, 3^{+}, 4^{-} )=0 $, hence there is no photon bending produced by this interaction.   

\subsection{Graviton bending}
\label{sec:graviton}
\begin{figure}[htbp]
\begin{center}
\begin{tikzpicture}[line width=1.5 pt,scale=.9]
	\draw[double, vector, line width = .75] (0:1.6) -- (-30:3);
	\draw[double, vector, line width = .75] (0:1.6) -- (30:3);
	\draw[double, fermionnoarrow, line width = .75] (180:1.6) -- (150:3);
	\draw[double, fermionnoarrow, line width = .75] (180:1.6) -- (210:3);
	\draw  [red, dashed, line width = 1.0] (90:1.5) -- (270:1.5);
	\draw[fill=white] (180:1.6) circle (.6);
	\draw[fill=white] (0:1.6) circle (.6);
	\draw[double, vector, line width = .75] (-1.08,.25) arc (120:60:2.16);
	\draw[->, line width = .5] (-.35,1.0) arc (100:80:2.16);
	\draw[->, line width = .5] (-.35,-1.0) arc (-100:-80:2.16);
	\draw[double, vector, line width = .75] (1.08,-.25) arc (-60:-120:2.16);
        \node at (179:1.6) {EH};
        \node at (0:1.6) {\large $R^3$};
        \node at (215:3.4) {$1^{\phi_{m}}$};
        \node at (145:3.4) {$2^{\phi_{m}}$};
%         \node at (34:3.4) {$p_3$};
%          \node at (-34:3.4) {$p_4$};
                   \node at (34:3.4) {$3^{++}$};
          \node at (-34:3.4) {$4^{--}$};
           \node at (120:1.15) {$\ell_1$};
          \node at (-120:1.15) {$\ell_2$};
          \node at (90:1.6) {$\pm \ \ \ \mp$};
          \node at (-90:1.6) {$\pm \ \ \ \mp$};
\end{tikzpicture}
\caption{The cut diagram contributing to the leading $R^3$ correction to gravitational scattering of a graviton (double wavy lines) off a massive scalar (double lines).}
\label{graviton-bending-fig}
\end{center}
\end{figure}
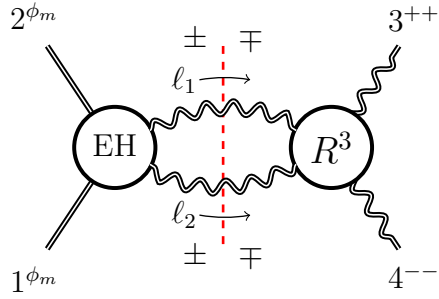

The relevant cut diagram is depicted in Figure \ref{graviton-bending-fig}.
The tree-level amplitudes entering this cut  are given by
\beq
A( 1^{\phi_m}, 2^{\phi_m}, \ell_1^{--}, \ell_2^{--} ) \ = \ - \left({\kappa\over 2}\right)^2 \, m^4 {\langle \ell_1 \, \ell_2\rangle^2 \over [ \ell_1\, \ell_2]^2}\Big[ {i \over (\ell_1 + p_1)^2 - m^2}\, + \, {i \over (\ell_1 + p_2)^2 - m^2}\Big]\ , 
\eeq
while  the amplitude in the $R^3$-deformed theory with two scalars and two gravitons is
 \cite{Broedel:2012rc}
 \beq
 A_{R^3} ( -\ell_2^{++}, -\ell_1^{++}, 3^{++}, 4^{--} ) \ = \ - i \left( {\kappa\over 2}\right)^2 \Big( {\alpha^\prime \over 4} \Big)^2
 ( \langle 4\, \ell_2\rangle [\ell_2\, 3] \langle 3\, 4\rangle)^2 { [ \ell_2\, \ell_1] [\ell_1\, 3] [ 3\, \ell_2] \over 
\langle \ell_2\, \ell_1\rangle \langle \ell_1\, 3\rangle \langle3\, \ell_2\rangle} \ . 
 \eeq
Using these ingredients, one quickly arrives at the following form for the $s$-cut:
\beq
\left. \mathcal{I}^{(1)}_{h} \right|_{s\textrm{-cut}} \ = \  -  \mathcal{D} \, m^4 \langle 4 | \ell_1 | 3]^4
\, 
\Big[ {1 \over (\ell_1 + p_1)^2 - m^2}\, + \, {1 \over (\ell_1 + p_2)^2 - m^2}\Big]\,
\Big[ {1 \over (\ell_1 - p_3)^2}\, + \, {1 \over (\ell_1 - p_4)^2}\Big]\ ,
\eeq
corresponding to four box topologies. Using \eqref{identity} 
we can recast this as 
\begin{align}
\begin{split} 
\mathcal{I}^{(1)}_{h} & = \   \mathcal{D}  \left({2\, m \,  \over \langle 3 | 2 | 4] }\right)^4 
\Big[ L^4 + E^4 + 6 L^2 E^2\Big] 
\\
&\Big[ {1 \over (\ell_1 + p_1)^2 - m^2}\, + \, {1\over (\ell_1 + p_2)^2 - m^2}\Big]\,
\Big[ {1\over (\ell_1 - p_3)^2}\, + \, {1\over (\ell_1 - p_4)^2}\Big] \frac{1}{\ell_1^2} \frac{1}{\ell_2^2}
\, . 
\end{split}
\end{align}
Following similar steps as in the previous case, and in particular keeping only the leading terms in the  limit \eqref{bendlim} we arrive at the result for the one-loop amplitude
 \beq
\label{gravitonbending}
A^{(1)}_h  =  \mathcal{D} \, N_h \, \left[ (m s)^4 \big(I_4 (s, t;m) + I_4 (s, u;m) \big)  + (m^2 s\, \omega )^2  I_3 (s;m) + \frac{3}{2} (m s\, \omega )^2    I_2 (s) \right]  , 
\eeq
where 
\begin{equation} 
\label{enh}
N_h := \big[ (2 m \omega) / \langle 3| 2| 4]\big]^4= N_\gamma^2\to 1
\end{equation}
 in the limit \eqref{bendlim} (see \eqref{ngamma}).  

A few comments on this result are in order.
    \begin{itemize} 
\item[ {\bf 1.}]
Compared to the graviton bending result in Einstein gravity \cite{Chi:2019owc}, the triangle and bubble contributions are suppressed by a factor of $s^2$, as expected from having four more derivatives compared to the Einstein-Hilbert action.  
 
\item[{\bf 2.}]
The box contribution  $I_4 (s, t;m) + I_4 (s, u;m)$ is purely imaginary (see \eqref{integrals}) and also appears (with a different coefficient) in the corresponding computation in the Einstein-Hilbert case  \cite{Bjerrum-Bohr:2014zsa, Chi:2019owc}. It  contributes an overall phase to the amplitude, and therefore will be dropped. 
 
\item[{\bf 3.}]
The result of the integral reduction, once we drop the box term, is exactly the same as we found for the scalar and photon case in \eqref{scalarbending} and \eqref{photonbending}. 
  
\item[{\bf 4.}]
We also note that since all four-point graviton amplitudes  do not receive contribution from the  $G_3$ interaction  \cite{vanNieuwenhuizen:1976vb,Broedel:2012rc}, graviton bending is not affected by this interaction. 
\end{itemize}

\subsection{From the amplitude to the potential and the bending angle}

\label{sec:bending}

Next we derive the potential, from which we can infer the bending angle. The potential is defined as in \eqref{potential}, where now, using \eqref{enbend}, we have $4 E_1 E_4 \to 4 m \omega$. As in \eqref{potentialbis} we decompose the potential into its classical and quantum contributions in momentum space: 
\beq
v_{\rm cl}  +\hbar v_{\rm qu} \ = \ \mathcal{D} {m^2 \omega\over 128} {s^2 \over \sqrt{-s}} \, + \, \hbar\, \mathcal{D} { m \omega\over 32 \pi^2}
s^2 \log(- s)  
\ . 
\eeq
Performing the Fourier transforms using the results in Appendix \ref{appen} we get 
\beq
V_{\rm cl}(\vec{r},\vec{p}) = ({\alpha^\prime} G_N)^2  {3 m^2 \omega \over 8 r^6}  \ , 
\qquad 
V_{\rm qu}(\vec{r},\vec{p}) =- ({\alpha^\prime} G_N)^2 {15 m \omega \over 2 \pi r^7}
\ . 
\eeq
The bending angle can then  be computed using the semiclassical formula  \cite{Donoghue:1986ya}
\begin{align}
\label{textbend}
\theta = - {b\over \omega} \int_{-\infty}^{+\infty}\! du \ {V^\prime ( b \sqrt{1 + u^2} ) \over \sqrt{ 1 + u^2}} 
\ , 
\end{align}
where $b$ is the impact parameter, 
with the result 
\beq
\label{finalbend}
\theta = (\alpha^{\prime} G_N)^2 \ {3  \over 64} 
\Big(15 \pi \,  {m^2\over b^6} \,\, - \, 
\hbar{1024 \over \pi} \, {m\over b^7}\Big)
 \ . 
 \eeq
 We can compare this result to that  obtained for scalars and photons \cite{Bjerrum-Bohr:2014zsa}, and gravitons  \cite{Chi:2019owc} in Einstein gravity. In those cases, the classical contribution is universal, as expected as a consequence of the  equivalence principle, but the quantum contribution differs for different particles.
 In our case,  both classical and quantum contributions are independent of the particle considered, and \eqref{finalbend} is the bending angle for scalar, photon and gravitons. It should be noted that the universality of the one-loop quantum correction is unexpected, and would clearly be interesting to confirm or disprove it  by higher-loop computations. We also note that our result for the bending angle is suppressed by a further factor of $1/b^4$ compared to the result of
  \cite{Bjerrum-Bohr:2014zsa,Chi:2019owc}, as expected from our use of a higher-derivative interaction. 

\subsection{Graviton bending in the bosonic string theory} 
\label{actionstring}

\begin{figure}[htbp]
\begin{center}
\begin{tikzpicture}[line width=1.5 pt,scale=.9]
	% \draw[fermion] (-30:1) -- (90:1) -- (210:1) -- cycle;
	%\draw (-45:1) -- (45:1);
	%\draw (45:1) -- (135:1);
	%\draw [double] (135:1) -- (225:1);
	%\draw  (225:1) -- (315:1);
	\draw[double, vector, line width = .75] (0:1.6) -- (-30:3);
	\draw[double, vector, line width = .75] (0:1.6) -- (30:3);
	%\draw[double, fermion] (135:1.4142) -- (135:3);
	%\draw[double, fermionnoarrow] (-1.8,.5) -- (150:3);
	\draw[double, fermionnoarrow, line width = .75] (180:1.6) -- (150:3);
	%\draw[double, fermionnoarrow] (-1.8,-.5) -- (210:3);
	\draw[double, fermionnoarrow, line width = .75] (180:1.6) -- (210:3);
	\draw  [red, dashed, line width = 1.0] (90:1.5) -- (270:1.5);
	\draw[fill=white] (180:1.6) circle (.6);
	\draw[fill=white] (0:1.6) circle (.6);
	\draw[double, vector, line width = .75] (-1.08,.25) arc (120:60:2.16);
	\draw[->, line width = .5] (-.35,1.0) arc (100:80:2.16);
	\draw[->, line width = .5] (-.35,-1.0) arc (-100:-80:2.16);
	\draw[double, vector, line width = .75] (1.08,-.25) arc (-60:-120:2.16);
        \node at (179:1.6) {EH};
        \node at (0:1.6) {\large $\Phi R^2$};
        \node at (215:3.4) {$1^{\phi_{m}}$};
        \node at (145:3.4) {$2^{\phi_{m}}$};
%         \node at (34:3.4) {$p_3$};
%          \node at (-34:3.4) {$p_4$};
          \node at (34:3.4) {$3^{++}$};
          \node at (-34:3.4) {$4^{--}$};
           \node at (120:1.15) {$\ell_1$};
          \node at (-120:1.15) {$\ell_2$};
          \node at (90:1.6) {$\pm \ \ \ \mp$};
          \node at (-90:1.6) {$\mp \ \ \ \pm$};
\end{tikzpicture}
\caption{The cut diagram contributing to the leading $(\Phi R^2)^2$ correction to gravitational scattering of a graviton (double wavy lines) off a massive scalar (double lines).}
\label{graviton-bending-fig2}
\end{center}
\end{figure}
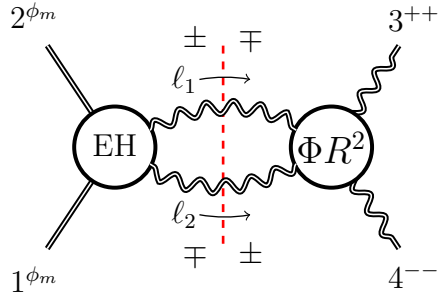

The modified EH action \eqref{action} that we considered is known to be contained in the low-energy effective action of the bosonic string theory \cite{Metsaev:1986yb}
\beq
\label{action-long}
S_B \ = \ -{2\over \kappa^2} 
\int\!d^4x \sqrt{ -g}  \Big[ R - 2 (\partial\Phi)^2 - \frac{1}{12} |dB|^2 + 
{\alpha^\prime\over 4} e^{ - 2 \Phi} G_2 + 
\alpha^{\prime \, 2} e^{- 4 \Phi} \Big(  {1\over 48} I_1 +{1 \over 24} G_3 \Big) + \mathcal{O} (\alpha^{\prime \, 3}) \Big] \ .
\eeq
In the definition of $S_B$ we have introduced
the Gauss-Bonnet combination $G_2=R^{\alpha \beta \mu \nu}R_{\alpha \beta \mu \nu}-4
R^{\alpha \beta}R_{\alpha \beta}+R^2$, $I_1 = {R^{\alpha \beta}}_{\mu \nu} {R^{\mu \nu}}_{\rho \sigma} {R^{\rho \sigma}}_{\alpha \beta}$ and $G_3 = I_1 - 2 {R^{\mu \nu \alpha}}_\beta {R^{\beta \gamma}}_{\nu \sigma} {R^\sigma}_{\mu \gamma \alpha}$.

A natural question is whether the additional terms in the full effective action of the bosonic string modify the computations presented so far in this paper.
The extra terms do not introduce  modifications of the three-graviton interaction \cite{vanNieuwenhuizen:1976vb,Broedel:2012rc}, and do not affect the three- and four-point graviton amplitudes. However, the ${R^{\mu \nu \alpha}}_\beta {R^{\beta \gamma}}_{\nu \sigma} {R^\sigma}_{\mu \gamma \alpha}$ term modifies the scalar potential, as shown recently in
\cite{Emond:2019crr} and discussed at the end of Section \ref{sec:potential}. 

In this section we  focus on the corrections to the graviton bending arising from the $G_2$ term. Here, a novel four-graviton amplitude with two positive and two negative helicity gravitons is produced due to two insertions of the $\Phi R^2$ contained in the $e^{-2 \Phi} G_2$ term of  \eqref{action-long}. Note that the $R^3$ term cannot produce a four graviton amplitude with this helicity configuration.

The  cut to consider is displayed in Figure \ref{graviton-bending-fig2}. 
The relevant amplitudes here are 
\beq
A ( 1^{\phi_m}, 2^{\phi_m}, \ell_1^{++} , \ell_2^{--}) \ = \ - \left({\kappa\over 2} \right)^2 { \langle \ell_2 | 2 | \ell_1]^4 \over s_{12}^2} \, 
\Big[ {i \over (\ell_1 + p_1)^2 - m^2} + { i \over (\ell_1 + p_2)^2 - m^2} \Big] 
\ , 
\eeq
while the $\Phi R^2$ amplitude is given by the simple expression \cite{Broedel:2012rc}
\beq
\label{4ptdila}
A_{\Phi R^2}(-\ell_1^{--}, 3^{++}, 4^{--}, -\ell_2^{++})  \ = \  - \left( {\kappa\over 2} \right)^2 \Big({\alpha^\prime \over 4}\Big)^2 {2 i \over (\ell_1 - p_4)^2} \langle \ell_1 \, 4 \rangle^4 [3\, \ell_2]^4 
\ ,
\eeq
which arises from two $\Phi R^2$-vertex insertions joined by a dilaton propagator.
The one-loop integrand compatible with the $s$-channel cut becomes 
\begin{align}
\begin{split}
\label{rrf}
\mathcal{I}^{(1)}_{h} & = 
\mathcal{D} {2\over s_{12}^2} \left({ 4\over \langle 3 | 2 | 4]  } \right)^4 
\Big[ L^4 + 6 L^2 \big[ p_2 \cdot (\ell_1 - p_4) \big]^2 E^2 + 
\big[ p_2 \cdot (\ell_1 - p_4) \big]^4 E^4 \Big] 
\\
&\cdot 
\Big[ {1 \over (\ell_1 + p_1)^2 - m^2} + { 1 \over (\ell_1 + p_2)^2 - m^2} \Big] \frac{1}{\ell_1^2} \frac{1}{\ell_2^2}
\ , 
\end{split}
\end{align}
where 
\begin{align}
\begin{split} 
L & := \ ( p_2 \ell_1) \big[ (p_2p_3) ( p_4 \ell_1) - (p_3p_4) (\ell_1 p_2) + (p_3 \ell_1) ( p_2p_4) \big] 
\\ &
+ (p_2p_4) \big[ (p_2p_3) (\ell_1 p_4) - (p_3 \ell_1 ) ( p_2p_4) + ( p_3 p_4) (\ell_1 p_2) \big] 
- m^2 (p_3p_4) (\ell_1 p_4) 
\ , 
\\
E^2 & := - \big[ \eps (p_2 p_3 p_4 \ell_1)\big]^2 = {\rm det}\, N \ , 
\end{split}
\end{align}
and $N$ is the matrix whose entries are the scalar products of the momenta within the Levi-Civita symbol. 
Performing the reductions, and taking the limit  \eqref{bendlim},  we obtain 
\begin{align}
\begin{split}
\label{gravitonbendingR2phi}
A^{(1)}_h =  \mathcal{D} \, N_h \, \Big[ & (4\,  m^2 \omega^2  s)^2 \big(I_4 (s, t;m) + I_4 (s, u;m) \big)  \, -
\, 
35  (m^2 s\omega )^2  I_3 (s;m) \\
&
 + 28 \,  ( m s\omega ) ^2 \, s\, I_3 (s) \ + \ 
 ( m s\omega ) ^2 
 \Big( - {251\over 6} + {3587\over 90} \eps\Big)
 I_2 (s) \Big] \ , 
\end{split}
\end{align}
where $N_h\to 1$ in the low-energy limit  (see \eqref{enh}).

Finally we compute the bending angle, following the same steps as in Section \ref{sec:bending}.  As before, we first compute the potential, from which we will then obtain the bending. The potential is defined in \eqref{potential}, where again,  using \eqref{enbend}, we have $4 E_1 E_4 \to 4 m \omega$. 
It can be  decomposed  into a  classical and quantum contribution in momentum space: 
\beq
v_{\rm cl}  +\hbar v_{\rm qu} \ = \ - \mathcal{D} {35 \, m^2 \omega \over 128} \,  {s^2 \over \sqrt{-s}} \, - \, \hbar\, \mathcal{D} \Big[ { 89 \, m \omega\over 96 \pi^2}
s^2 \log(- s)  + {7\, m\omega  \over 32 \pi^2}  \, s^2  \log^2 (-s) \Big]
\ .
\eeq
%where we have also included a factor of two from the two possible internal helicity configurations. 
Performing the Fourier transforms using results in Appendix \ref{appen} and reinstating couplings and  the appropriate kinematic  prefactor, we arrive at
\begin{align}
\begin{split}
V_{\rm cl}(\vec{r},\vec{p}) &=- ({\alpha^\prime}G_N)^2  \, {105  \over 8} {m^2 \omega \over r^6} \ , 
\\
V_{\rm qu}(\vec{r},\vec{p}) &= ({\alpha^\prime}G_N)^2 \, {m \omega \over r^7} \, \Big[
  %{
  %890 
  {702\over \pi} \,  - {210\over  \pi } \log (r/r_0
  %\mu e^{\gamma_E}
  )  \Big]
\ . 
\end{split}
\end{align}
Using again \eqref{textbend}, we arrive at the final result for the bending angle in the presence of a $\Phi R^2$ coupling: 
\beq
\label{finalbendbis}
\theta =(\alpha^{\prime} G_N)^2 
\Big\{ -\frac{1575 \, \pi  
   }{64}{m^2\over b^6} +\hbar {64\over \pi} 
   \Big[-21 
   \log
   \big(b/ (2r_0)\big)+{229\over 4}
    \Big]{m \over b^7} 
\Big\}
\ 
 \ .
 \eeq
 It is interesting to compare \eqref{finalbendbis} with \eqref{finalbend}. We note that the classical contributions to  these two angles have opposite signs, and the $\Phi R^2$ contribution is larger than the $R^3$ contribution by a factor of 35.
 Similar comments apply to the quantum correction. Hence in the bosonic string the combined bending angle would be dominated by the $\Phi R^2$ contribution. 
 
Finally, we briefly consider what would happen to the bending angle if  the dilaton acquires a mass $M_{\phi}$, as expected in phenomenologically realistic models where the dilaton is stabilised. The main modification occurs in the four-graviton amplitude \eqref{4ptdila}, which now would be derived by joining two $R^2\phi$ vertices with a massive dilaton propagator, thus replacing $(\ell_1 - p_4)^2$ with $(\ell_1 - p_4)^2 - M_{\phi}^2$. As a first approximation, we can consider the dilaton as very heavy and thus replace its propagator with $-1/M_{\phi}^2$. 
 Following steps identical to those in the massless case, one arrives at the following expression for the bending angle: 
 \beq
\label{finalbendter}
\theta =(\alpha^{\prime} G_N)^2 {\omega^2\over M_{\phi}^2} 
\Big[\frac{1575 \, \pi  
   }{64}{m^2\over b^6}\,  -\, \hbar {1536\over \pi} 
    {m \over b^7} 
\Big]
\ 
 \ , 
 \eeq
 which has a large suppression factor arising from the $(\omega / M_{\phi})^2$ prefactor compared to the bending angle \eqref{finalbendbis} for the case of a massless dilaton. 

\section{Closing comments}
\label{sec:end}

We wish to conclude  with a summary of some open problems and possible future directions of our work, which
clearly only touches on the  tip of an iceberg of possible higher-derivative modifications that can be contemplated.
\begin{itemize}
\item[{\bf 1.}]~It would be interesting to consider particles coupled non-minimally to the graviton {\it e.g.} the photon coupled
to the Riemann tensor as $\alpha_\gamma \int d^4x \sqrt{-g} F^{\mu\nu} F^{\alpha\beta} R_{\mu\nu\alpha\beta}$. The leading
correction to the amplitude would then come from a single graviton-exchange diagram.
\item[{\bf 2.}]~It would be interesting to understand the universality ({\it i.e.}~spin-independence) of the quantum corrections to the
particle bending. In pure gravity only the classical corrections are universal in consonance with the equivalence principle.
\item[{\bf 3.}]~Can $\alpha^\prime$ be made large enough, and  consistent with known constraints, to produce effects that are comparable
with PN$x$ correction from pure gravity, and for what $x$? 
\end{itemize}

%\vspace{1cm}

\section*{Acknowledgements}

We would like to thank Manuel Accettulli Huber, David Berman,  Stefano De Angelis, Rodolfo Russo  and Chris White for  discussions, and Andrea Cristofoli for detailed discussions on the post-Minkowskian expansion in gravity. 
This work  was supported by the Science and Technology Facilities Council (STFC) Consolidated Grant ST/P000754/1 
\textit{``String theory, gauge theory \& duality"}, 
and by the 
European Union's Horizon 2020 research and innovation programme under the Marie Sk\l{}odowska-Curie grant agreement 
No.~764850 {\it ``\href{https://sagex.org}{SAGEX}"}.

\pagebreak

\appendix
\section{Integrals and Fourier transforms}
\label{appen}

The  expression for the integral functions occurring in our calculations, expanded up to the relevant orders in $\eps$, and keeping only terms with an $s$-channel discontinuity, are: 
\begin{align}
\begin{split}
\label{integrals}
I_2(s) & = \  i c_\Gamma {( -s)^{- \eps}\over \eps ( 1 - 2 \eps)} \ \simeq \ {i\over 16 \pi^2} \Big[ {1\over \eps} - \log ( - s)  \Big]\ , \\
I_3 (s) & =  -i c_\Gamma {(-s)^{-1-\eps}\over \eps^2}\ \simeq \ 
{i\over 16 \pi^2} {1\over s} \Big[ {1\over \eps^2} - {\log ( -s) \over \eps} + {1\over 2} \log^2 ( -s ) \Big]\, , 
\\
I_3 (s; m) & = - {i\over 32} \Big[ {1\over m  \sqrt{ -s} } + { \log ( - {s/m^2} ) \over \pi^2 m^2} \Big] + \cO (\sqrt{s} )\ , \\
I_4(s, t; m) + I_4 (s, u; m) & \simeq \  {i\over 16 \pi \, s \, ( m \omega)} \cdot i \, \Big[  {1\over \eps} 
\, -  \, 
\log \left(-\frac{s}{m^2}\right)\Big]\ , 
\end{split}
\end{align}
where 
\beq
c_\Gamma = {\Gamma(1+\eps) \Gamma^2 (1-\eps) \over (4 \pi)^{2 - \eps} \Gamma (1 - 2 \eps)}\, ,
\eeq 
and $f(\eps)$ is a kinematic-independent function that will contribute to any of the physical quantities computed in this paper as it gives rise to terms that vanish when Fourier transformed. 
We also quote the relevant Fourier transforms  used in the text: 
\beq
\label{a3}
\int\!{d^d q \over (2 \pi)^d} \ e^{ i \vec{q} \cdot \vec{r}} \,  |\vec{q}\, |^\alpha \ = 
\left({2\over r}\right)^{d+\alpha} {\Gamma\left( {d + \alpha\over 2}\right) \over (4 \pi)^{d/2} \Gamma\left( - {\alpha \over 2} \right)}\ , 
\eeq
as well as  
\beq
\label{a4}
\int\!{d^3 q \over (2 \pi)^3} \ e^{ i \vec{q} \cdot \vec{r}}   |\vec{q}\, |^4 \log ( q^2) \ = \
-\dfrac{60}{\pi} \dfrac{1}{r^7}\ ,
\eeq
and
\begin{align}
\begin{split}
&\int\!{d^3 q \over (2 \pi)^3} \ e^{ i \vec{q} \cdot \vec{r}}  \,  |\vec{q}\, |^4 \log^2 \left({q^2\over \mu^2}\right) \ = \ 
{4\over \pi } {1\over r^7}\big[ 60 \log (r/r_0)  - 137\big]
   \ , 
   \end{split}
\end{align}
where $r_0:= (\mu e^{\gamma_E})^{-1}$.

\pagebreak

	\bibliographystyle{utphys}
	\bibliography{remainder}

\end{document}